\documentclass[%
 reprint,
 amsmath,amssymb,
 aps,prl
]{revtex4-1}

\usepackage{graphicx}
\usepackage{dcolumn}
\usepackage{bm}

\usepackage{color}  

\begin{document}

\preprint{APS/123-QED}

\title{Selective coupling of optical energy into the fundamental diffusion mode of a scattering medium
}

\author{Oluwafemi S. Ojambati, Hasan Y{\i}lmaz, Ad Lagendijk, Allard P. Mosk, and Willem L. Vos } 
\affiliation{
Complex Photonic Systems (COPS), MESA+ Institute for Nanotechnology, University of Twente, P.O. Box 217, 7500 AE Enschede, The Netherlands
}

\date{June 17th, 2015}

\begin{abstract}
We demonstrate experimentally that optical wavefront shaping selectively couples light into the fundamental diffusion mode of a scattering medium. 
The total energy density inside a scattering medium of zinc oxide (ZnO) nanoparticles was probed by measuring the emitted fluorescent power of spheres that were randomly positioned inside the medium. 
The fluorescent power of an optimized incident wavefront is observed to be enhanced compared to a non-optimized incident wavefront. 
The observed enhancement increases with sample thickness. 
Based on diffusion theory, we derive a model wherein the distribution of energy density of wavefront-shaped light is described by the fundamental diffusion mode.
The agreement between our model and the data is striking, not in the least since there are no adjustable parameters.
Enhanced total energy density is crucial to increase the efficiency of white LEDs, solar cells, and of random lasers, as well as to realize controlled illumination in biomedical optics.
\end{abstract}

\pacs{Valid PACS appear here}
\maketitle
%
%
%

\begin{figure}[tbp]
\center
\includegraphics[scale = 1]{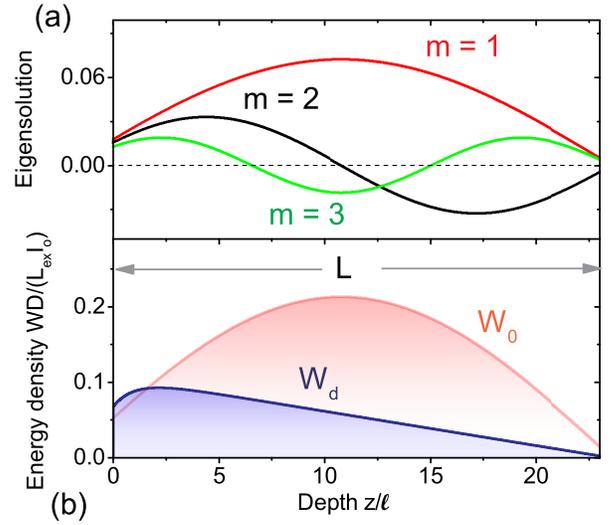}
\caption{(Color) (a) The first three eigensolutions of the diffusion equation, where $m$ is index of the eigensolution. 
(b) The energy density of optimized light and non-optimized light are shown as the red and blue curves respectively. 
The energy density is reduced with the diffusion constant $D$, the incident intensity $I_{\rm{o}}$ and the effective thickness of the sample $L_{\rm{ex}} = L + z_{\rm{e1}} + z_{\rm{e2}}$, $z_{\rm{e1}}$ and $z_{\rm{e2}}$ are the extrapolation lengths at the front and back surfaces of the sample respectively. }
\label{fig:figure1}
\end{figure}

Numerous physical phenomena are described by diffusion~\cite{fick1854PoggAnnPhysChem, einstein1905AnnPhys, ishimaru1978Book, blaaderen1992JChemPhys, philibert2006DiffFund, akkermans2007book, bonnell2012RevModPhys, pierrat2014PNAS}. 
Diffusion is a process that leads to uniform spreading of matter or energy as a result of randomness~\cite{ishimaru1978Book, glicksman1999diffusion}. 
Diffusion theory accurately describes the propagation of the energy density of multiply scattered waves in disordered scattering media~\cite{ishimaru1978Book, pierrat2014PNAS, Page1995PRE, wiersma1996PRE, vanRossum1999RevModPhys, scheffold1999Nature, yamilov2014PRL}.
Upon averaging over the disorder, waves become diffuse after a distance of the order of one transport mean free path $\ell $ and the energy density of the waves acquires a typical shape, shown in Fig.~\ref{fig:figure1}(b).
The derivative of the energy density at the exit surface is related to the transport of energy, and yields the waves-equivalent of the well-known Ohm's law, $T \approx \ell /L $, where $L$ is thickness of the scattering medium. 
 
In a slab geometry, the solution of the diffusion equation can be expressed as a sum over a complete set of eigensolutions with imaginary frequency~\cite{carslawBook1959}.
In Fig.~\ref{fig:figure1}(a), we show the first three eigensolutions.
When a plane wave is incident on a scattering medium, energy is coupled into all eigensolutions, which sums up to give the non-optimized energy density $W_{\rm{d}}$ shown in Fig.~\ref{fig:figure1}(b). 
A fundamental question we seek to address is the opportunity 
of changing the internal energy by selectively coupling energy only into the fundamental diffusion eigenmode with index $m = 1$ shown in Fig.~\ref{fig:figure1}(a).
It is of particular interest when the total energy coupled into the fundamental diffusion mode is greater that of the  non-optimized energy density $W_{\rm{d}}$ as shown in Fig.~\ref{fig:figure1}(b).
In some particular cases, the fundamental diffusion eigenmode has a greater total energy density than the unoptimized energy density $W_{\rm{d}}$ as shown in Fig.~\ref{fig:figure1}(b).
In the case of light, which is the subject of our work, an enhanced energy density inside the scattering medium is important for applications, such as enhanced energy conversion in white LEDs \cite{krames2007JDispTechnol, phillips2007LaserPhotonRev, ogi2013ECIJSolidStateSciTechnol, Leung2014OptExp}, efficient light harvesting in solar cells \cite{levitt1977ApplOpt, polman2012Nature, isabella2010ApplPhysLett}, low-threshold random lasers \cite{lawandy1994Nature, genack1994Nature, wiersma1996PRE}, and controlled illumination in biomedical optics \cite{yizhar2011Neuron}.

The total transmitted intensity through a scattering medium can be made to differ from Ohm's law by wavefront shaping \cite{freund1990PhysicaA,vellekoop2007OptLett,vellekoop2008PRL, mosk2012NatPhoton, aulbach2011PRL, popoff2014PRL}, time reversal \cite{lerosey2004PRL, lerosey2007Science},  phase conjugation~\cite{leith1966JOptSocAm, yaqoob2008NatPhoton}, and control based on transmission matrix~\cite{kim2012NatPhoton, popoff2010PRL}. 
In wavefront shaping, the spatial phase of the incident field on the scattering medium is controlled in order to enhance the intensity in a diffraction-limited spot at the back surface of the sample. 
Only numerical calculations~\cite{choi2011PRB, davy2015arx, liew2015arx} and a single-realization experiment of elastic waves~\cite{aubry2014PRL} with a shaped incident wavefront have been used to study the distribution of energy density inside a two-dimensional (2D) scattering medium.  
The distribution observed in these calculations and in the single-realization experiment is a symmetric function peaked at the middle of the sample, which is similar to the fundamental eigenmode $m = 1$. 
The change in the energy density has so far not been experimentally observed inside a three-dimensional (3D) scattering medium. 

In this Letter, we demonstrate experimentally the selective coupling of light into the fundamental mode of the diffusion equation by using wavefront shaping. 
We probe the total internal energy, which is integral of the position-dependent energy density inside a 3D scattering medium. 
As a probe, we employ fluorescent spheres randomly positioned inside the medium. 
We observe that the total energy increases when the incident light is shaped. 
The enhancement in fluorescent power increases with sample thickness. 
To interpret our results, we propose a model wherein the energy density of wavefront-shaped light is described by the fundamental eigensolution of the diffusion equation. 
Our model has no adjustable parameters and agrees well with the experimental results. 

\begin{figure}[tbp]
\center
\includegraphics[width=0.45\textwidth]{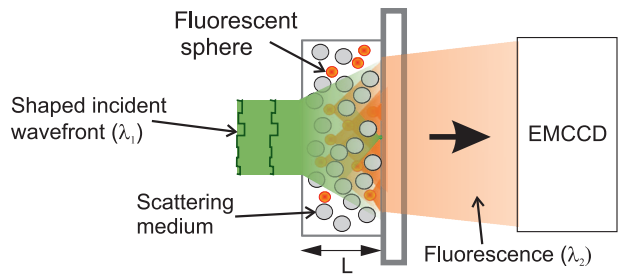}
   \caption{(Color) Schematic drawing of the method to probe the total energy density inside a scattering medium. 
The scattering medium is an ensemble of disordered ZnO particles in air. 
The medium is illuminated with a shaped incident wavefront such that the incident light at $\lambda_1$ (green intensity) is optimized on a diffraction-limited spot at the back of the sample. 
The scattering medium is lightly doped with fluorescent spheres randomly positioned inside the medium to probe the energy density inside the sample. 
The total fluorescent power emitted from the fluorescent spheres at $\lambda_2$ (red intensity) is measured by EMCCD.}
    \label{fig:figure2}
  \end{figure}

In our experiments, we study a scattering medium, which is a layer of spray-painted zinc oxide (ZnO) nanoparticles on a microscope glass slide of thickness $0.17 \, \rm{mm}$.  
The transport mean free path of similar samples was determined from total transmission measurements to be 0.6 $\pm $ $0.2 \,\rm{\mu m}$ \cite{elbertThesis}. 
Inside the ZnO samples, dye-doped polystyrene spheres with diameter 50 $\rm{nm}$ are randomly dispersed, as illustrated in Fig.~\ref{fig:figure2}.
The fluorescent spheres are excited by incident laser light with wavelength $\lambda_1 = 561\, \rm{nm}$ and emit fluorescent light at a different wavelength $\lambda_2 = 612\, \rm{nm}$.
In order to ensure that the spatial distribution of the energy density at $\lambda_1$ inside the scattering medium is not perturbed by the absorption from the probing fluorescent spheres~\cite{liew2014PRB}, we use samples with a low density of spheres and with a high albedo~\cite{sup_material}. 
  
A phase-only liquid crystal spatial light modulator (SLM) (Holoeye Pluto) shapes the wavefront of the laser light incident on the sample, such that the intensity is focused in a diffraction-limited spot at the back surface of the sample. 
We used the piece-wise sequential algorithm described in Ref. \cite{vellekoop2007OptLett} to find an optimized incident wavefront. 
The back surface of the sample is imaged to the chip of an electron multiplying charged-coupled device (EMCCD) camera to collect the total fluorescent intensity at $\lambda_2$. 
A combination of a dichroic mirror, a low-pass filter and a notch filter blocks the incident light at $\lambda_1$ from reaching the EMCCD~\cite{sup_material}. 

To obtain ensemble-averaged data that can be compared to theory, we need to average over different realizations of scatterers in the sample. 
We therefore performed automated sequences of wavefront shaping measurements while between two consecutive measurements, the sample was translated to a different realization by a piezo stage. 
In each measurement, we measured the total fluorescent power $P^{\rm{o}}_{\rm{f}}$ with the optimized pattern on the SLM and then measured the total fluorescent power $P^{\rm{n}}_{\rm{f}}$ with an incident wavefront optimized for a different uncorrelated position.
We define the fluorescent power enhancement $\eta_{\rm{f}}$ as the ratio of the two fluorescent powers, $\eta_{\rm{f}} \equiv P^{\rm{o}}_{\rm{f}}/P^{\rm{n}}_{\rm{f}}$.

We determine the fidelity $|\gamma|^2$ that quantifies the overlap of the experimentally generated field with the ideal controlled field \cite{vellekoop2008PRL}. 
The fidelity $|\gamma|^2$ achieved in a specific experimental run can be obtained by dividing the optimized power in the target spot by the average total transmitted power without optimization.
We performed 100 wavefront shaping experiments, each on a different position on a $L$ = 22.8 $\rm{\mu m}$ $\pm$ 0.95  thick sample. 
In each experiment, we determine the fidelity $|\gamma|^2$ and fluorescent power enhancement $\eta_{\rm{f}}$.
Factors such as inhomogeneity of the sample thickness, measurement noise and instability in environmental conditions result into variation of the fidelity $|\gamma|^2$~\cite{vellekoop2008OptComm, yilmaz2013BiomedOptExpress}. 
Although these factors are undesirable, they have the advantage of giving a wide range of $|\gamma|^2$ to investigate.

\begin{figure}[tbp]
\center
\includegraphics[scale = 1]{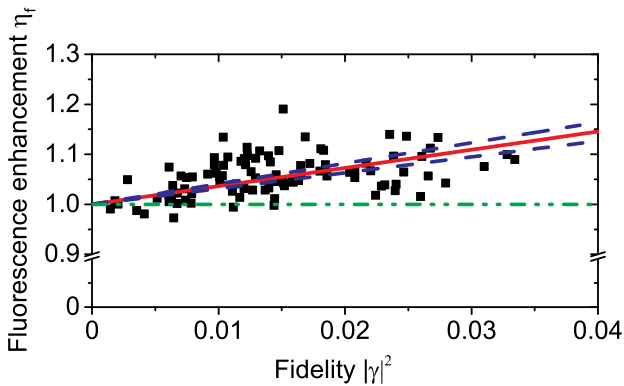}
\caption{(Color online) Measured fluorescent power enhancement $\eta_{\rm{f}}$ versus the fidelity $|\gamma|^2$  for an $L$ = 22.8 $\rm{\mu m}$ $\pm$ 0.95 thick ZnO sample. 
The black squares are 100 experimental data points obtained at different positions on the sample. 
The solid red curve is a linear regression through the data and the blue dashed curves are the 95$\%$ confidence interval. 
The green dash-dotted curve is the expected curve if the distribution of light with optimized wavefronts were the same as with diffuse light.}
\label{fig:figure3}
\end{figure}

In Fig.~\ref{fig:figure3} we show the enhancement in fluorescent power $\eta_{\rm{f}}$ versus the fidelity $|\gamma|^2$. 
Interestingly, we see an enhancement in the fluorescent power by up to about 10$\%$ as $|\gamma|^2$ increases to about 0.035. 
This increase implies that the total energy density for optimized incident wavefronts is higher than the total energy density of unoptimized incident wavefronts. 
If the spatial distribution of energy density would be unmodified by wavefront shaping, then the fluorescent intensity enhancement would have been constant at 1, which is obviously not the case. 
The slope of the linear regression fit to the data is 3.6 with a standard error of 0.2, and upper and lower 95$\%$  confidence intervals of 4.1 and 3.2, respectively. 
The correlation coefficient $r$ \cite{edwards1973Book} of the data is 0.9, which confirms that our data show a linear trend. 
The measured fluorescent power enhancement $\eta_{\rm{f}}$ has contributions from both the perfectly shaped wavefront and from the background intensity, which is the uncontrolled part of the intensity. 
We therefore express $\eta_{\rm{f}}$ in terms of the fidelity $|\gamma|^2$ as $\eta_{\rm{f}} = \eta^{\rm{e}}_{\rm{f}} \, |\gamma|^2 + (1 - |\gamma|^2)$, where $\eta^{\rm{e}}_{\rm{f}}  $ is the fluorescent power enhancement extrapolated to the limit of perfect fidelity $|\gamma|^2 \rightarrow 1$. 
The second term is the contribution from the background intensity. 
For the result shown in Fig.~\ref{fig:figure3}, we find $\eta^{\rm{e}}_{\rm{f}}  = 4.6 \pm 0.48 $.  

\begin{figure}[tbp]
\center
\includegraphics[scale = 1.1]{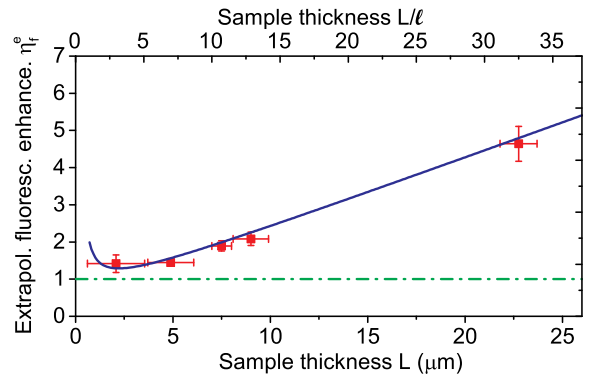}
\caption{(Color online) Fluorescent power enhancement $\eta^{\rm{e}}_{\rm{f}}$ in ZnO scattering samples versus sample thickness. 
The red squares are the measured fluorescent power enhancement extrapolated to unity intensity control. 
The blue solid line is the calculated fluorescent power enhancement from Eq.~\ref{eq:eta_f}. 
The green dash-dotted curve is for an invariant distribution of energy density along the sample depth.}
\label{fig:figure4}
\end{figure}

We studied samples with thicknesses $L$  ranging from  2 $\rm{\mu m}$ to 22 $\rm{\mu m} $ and on each sample we performed 100 to 130 wavefront shaping experiments.
Since the fidelity $|\gamma|^2$ decreases with increasing sample thickness~\cite{vellekoop2008PRL}, we derived for each sample the extrapolated fluorescent power enhancement $\eta^{\rm{e}}_{\rm{f}}$ to allow for a comparison between samples. 
In Fig.~\ref{fig:figure4}, we show that the extrapolated fluorescent power enhancement $\eta^{\rm{e}}_{\rm{f}}$ increases with sample thickness $L$, which means that wavefront shaping serves to optimally store energy in the volume of the medium. 
The uncertainty in the extrapolated fluorescent power enhancement increases with sample thickness, since the fidelity decreases for thicker samples. 
The horizontal error bars denote the standard deviation of the measurement of the sample thickness on different positions on the sample. 
For perfect fidelity, the total fluorescent power inside a 22.8 $\rm{\mu m}$ thick sample is 4.6 times greater than the total fluorescent power for non-optimized light. 

\begin{figure}[tp!]
\begin{center}
\includegraphics[scale = 1]{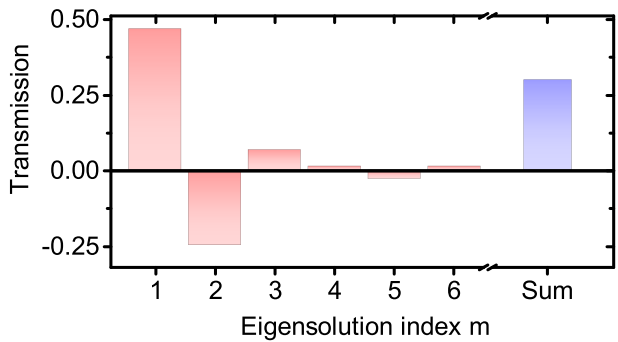}
\end{center}
\caption{(Color online) The contribution to the total transmission of the first six eigensolutions are represented by the red bars while the total transmission, which is the sum of all  transmissions of the individual eigensolutions is represented by the blue bar.}
\label{fig:figure5}
\end{figure}

To interpret our experimental results, we employ diffusion theory \cite{sup_material}. 
We obtained the diffuse energy density $W_{\rm{d}}$ shown in Fig.~\ref{fig:figure1}(b) from the diffusion equation. 
For light with an optimized incident wavefront, the distribution of the energy density $W_{\rm{o}}$ inside the medium is \emph{a-priori} unknown. 
With the optimized phase incident on the sample, light is coupled to the transmission eigenmodes of the wave equation with the highest transmission. 
Since both the wave equation and the diffusion equation describe the same physical system, we expect that the ensemble-averaged energy density of the transmission eigenmodes with the highest tranmission is equivalent to the diffusion eigensolution that contributes the most to the total transmission.
We show in Fig.~\ref{fig:figure5} the contribution to the total transmission \cite{footnote_Ttot} of the first six eigensolutions.
The fundamental eigensolution $m = 1$ contributes the most to the total transmission, even more than the total transmission, which is a summation of contribution of all the eigensolutions. 
We therefore hypothesize that the energy density distribution of optimized light is identical to the fundamental eigensolution of the diffusion equation. 
The validity of this hypothesis is verified when we compare our model to experimental data. 

It has been shown experimentally in Ref.~\cite{vellekoop2008PRL} and theoretically in Ref. \cite{beenakker1997RevModPhys} that the total transmission of optimized light is equal to $T_{\rm{o}} = 2/3$. 
We therefore scale the energy density of wavefront-shaped light such that the total transmission is equal to $T_{\rm{o}} = 2/3$. 
In Fig.~\ref{fig:figure1}(b), we show the scaled energy density of light with an optimized wavefront that clearly deviates from the distribution of diffuse light. 
In addition, Davy \emph{et al.} theoretically calculated the internal energy density distribution of transmission channels with a transmission coefficient of unity and found a parabolic solution~\cite{davy2015arx}.
Since the boundary conditions used in Ref.~\cite{davy2015arx} only apply to a medium that is index-matched to the surrounding media, the model does not pertain to our experiments. 
Nevertheless, it is interesting to note that the parabolic function found by Davy \emph{et al.} and our sine function are both symmetric functions peaked in the middle of the sample. 
 
Taking into account the diffusion of the emitted fluorescent light propagating through the sample we analytically model the fluorescence enhancement as
%
\begin{widetext} 
\begin{equation}
\eta^{\rm{e}}_{\rm{f}}(L) = 
\frac{2 L_{\text{ex}}^2 \sec (\frac{\pi  z_{\text{e2}}}{L_{\text{ex}}}) \Big{[} 
\pi  z_{\text{e1}} \cos (\frac{\pi  z_{\text{e1}}}{L_{\text{ex}}})
- L_{\text{ex}} [\sin (\frac{\pi z_{\text{e1}}}{L_{\text{ex}}}) - \sin (\frac{\pi  L^\prime}{ L_{\text{ex}}})]
- \pi L^\prime \cos (\frac{\pi  L^\prime)}{L_{\text{ex}}}) \Big{]}}
{3 \pi ^3 \Big{[} \frac{L z^\prime_{\text{inj}} [L^2
+ 3 L (z_{\text{e1}}
+ z_{\text{e2}})
+ 6 z_{\text{e1}} z_{\text{e2}}]}
{6 L_{\text{ex}}}
+z_{\text{inj}}^2 (L^\prime+z_{\text{inj}})e^{-\frac{L}{z_{\text{inj}}}} 
- z_{\text{inj}}^2 z^\prime_{\text{inj}}\Big{]}} \, ,
\label{eq:eta_f}
\end{equation}
\end{widetext} 
%
where $L_{\text{ex}} = L + z_{e1} + z_{e2} $ is the effective sample thickness, $L^\prime = L + z_{e1}$, and $z^\prime_{\text{inj}} = z_{\text{inj}} + z_{e1}$, and $z_{\text{inj}}$ is the injection depth at which the incident light becomes diffuse and it accounts for the angular distribution of the incident shaped wavefront \cite{deBoerThesis}.   
In order to compare our model to our experimental results, we plot in Fig.~\ref{fig:figure4} the analytic model for $\eta^{\rm{e}}_{\rm{f}}$ versus sample thickness $L$. 
Our model agrees well with our experimental result. 
There are no freely adjustable parameters in our model. 
If the spatial distribution of both wavefront-shaped and unwavefront-shaped light would have been the same, then $\eta_{\rm{f}}$ would been constant equal to 1 for all sample thicknesses as shown in Fig.~\ref{fig:figure4}, which does not agree at all with our observations. 
The excellent agreement between our model and our experimental results confirms the validity of our hypothesis that the distribution of wavefront-shaped light inside the medium is modified, and that energy has been coupled into the fundamental diffusion mode.

In our experiments we obtain the fluorescent power enhancement $\eta_f$ rather than the energy density enhancement $\eta_{\rm{ed}}$.
Therefore we define the enhancement of the energy density to be $\eta_{\rm{e}} \equiv W^\prime_{\rm{o}}/W^\prime_{\rm{d}}$, where $W^\prime_{\rm{o}}$ and $W^\prime_{\rm{d}}$ are the energy densities for optimized light and unoptimized light, respectively, both integrated over the whole sample thickness. 
We obtain 
\begin{equation}
\eta_{\rm{ed}}(L) = \frac{2}{3} \,\eta_{\rm{f}}(L) + O(L/l) \, ,
\label{eq:eta_flu_functionOf_eta_ed}
\end{equation} 
%
where $O(L/l)$ includes higher orders of the series expansion in terms of $L/l$, see supplementary material~\cite{sup_material}. 
We see from Eq.~\ref{eq:eta_flu_functionOf_eta_ed} that the total fluorescent power depends on the total energy density inside the medium.
Therefore the observed increase of the fluorescence is indeed a measure of the increase of the energy density.

In summary, we have experimentally demonstrated and theoretically modeled the selective coupling of light into the fundamental diffusion mode, by increasing the total energy density inside a scattering medium of ZnO nanoparticles by using wavefront shaping. 
Our results apply to other wave control methods in scattering media, such as time reversal, phase conjugation, and control based on transmission matrix as well as to other types of waves such as microwaves, acoustic waves, elastic waves, surface waves, and electron waves. 
We expect our results to be relevant for applications that require enhanced total optical energy density such as efficient light harvesting in solar cells especially in near infrared where silicon has low absorption; for enhanced energy conversion in white LEDs, which serves to reduce the quantity of expensive phosphor; for low threshold and higher output yield of random lasers; as well as in homogeneous excitation of probes in biological tissues.
Last but not least, it will be fruitful to investigate possible relationships between the fundamental diffusion eigensolution and the universal diffusion time obtained in Refs.~\cite{pierrat2014PNAS, blanco2003EuroPhysLett, tiggelen1993JPhysAMathGen}.

\begin{acknowledgements}
We thank Henri Thyrrestrup, Bas Goorden, Jin Lian, and Sergei Sokolov for useful discussions and Cornelis Harteveld  for technical assistance. 
This project is part of the research program of the ``Stichting voor Fundamenteel Onderzoek der Materie'' (FOM) ``Stirring of light!'', which is part of the ``Nederlandse Organisatie voor Wetenschappelijk Onderzoek'' (NWO), NWO-Vici, DARPA, and STW.
\end{acknowledgements}

\end{document}